\documentclass[aps,twocolumn,
showpacs,pra,amssymb, amsmath]{revtex4}
\usepackage{amsmath,latexsym,amssymb}
\usepackage[dvips]{graphicx}
\usepackage{amsmath}
\usepackage{latexsym}
\usepackage{amssymb}
\usepackage{bm}

\begin{document}

\title{Classification of mixed high-dimensional multiparticle systems}
\author{Koji Nagata}
\affiliation{
CREST Research Team for Interacting Carrier Electronics,
School of Advanced Sciences, The Graduate University for
Advanced Studies (SOKENDAI), Hayama, Kanagawa, 240-0193, Japan
}
\date{\today}

\begin{abstract}
We present an inequality that classifies mixed multipartite systems
of an arbitrary dimension with respect to separability 
and positivity of partial transpose properties.
This inequality gives a way to experimentally classify the observed state 
of multipartite systems
of an arbitrary dimension.
The inequality also implies that a sufficient condition 
for a density operator to have no positive partial 
transpose with respect to any subsystem is that the fidelity to 
a generalized Greenberger-Horne-Zeilinger 
state {[A. Cabello, Phys. Rev. A {\bf 63}, 022104 (2001)]} is larger than 1/2 for mixed multipartite systems
of an arbitrary dimension.
\end{abstract}

\maketitle

\section{Introduction }

In the quantum information processing, 
entanglement is a very important ingredient.
Moreover, understanding about the nature of entanglement
we can approach the core of a quantum world. 
Werner proposed the class of 
separable (classically correlated) states
and showed that there exists some inseparable state which can admit local hidden variable model for two-particle systems \cite{bib:Werner1}. 
Peres presented a necessary condition for separability, namely, the partial transpose of a density operator is a positive operator \cite{bib:Peres1}.
The partial transpose of an operator on 
Hilbert space ${\cal H}_1\otimes{\cal H}_2$ is defined by
\begin{eqnarray}
\left(\sum_l A^1_l\otimes A^2_l\right)^{T_1}=\sum_l{A_l^1}^{T} \otimes A_l^2,
\end{eqnarray}
where the superscript $T$ denotes transposition in the given basis.
Horodecki {\it et al}. showed that
the condition is also sufficient as to composite quantum systems on
${\cal H}={\cal C}^2\otimes {\cal C}^2$ and 
${\cal H}={\cal C}^2\otimes {\cal C}^3$
\cite{bib:Horodecki}. 
D\"ur {\it et al}. formulated criterions that determine to which 
class, with respect to separability and distillability properties, a density operator of multiqubit systems
belongs \cite{bib:Dur1,bib:Dur2}.
For mixed high-dimensional multiparticle systems, the criterions that determine whether a density operator is separable or not with 
respect to a subsystem become 
much more complicated (see, for example, Refs.~\cite{bib:Kraus,bib:Karnas}).
Furthermore, experiments in higher-dimensional systems (spin-1 systems) 
have been reported so far \cite{bib:Spin1,bib:Spin2}.
It will be helpful to give some formula 
that allows calculating 
the minimum of $k$ such that a given state is not $k$ separable
for mixed high-dimensional multiparticle systems.
And it will be useful to propose how to classify an observed state experimentally for such systems.

The purpose of this paper is to give an inequality that classifies a given state of multipartite systems
of an arbitrary dimension with respect to separability 
and positivity of partial transpose properties
by means of the expectation value of 
an operator, which is a linear combination of two projection operators.
This is possible by generalizing the result obtained by D\"ur {\it et al}.
Applying the inequality 
derived in this paper, we present a sufficient condition 
for a density operator to have no positive partial 
transpose with respect to any subsystem,
which is that the fidelity to a quantum state 
belonging to the generalized Greenberger-Horne-Zeilinger (GHZ) state class
that was discussed by Cabello \cite{bib:Cabello} 
is larger than 1/2,
for mixed multipartite systems of an arbitrary dimension.


\section{Inequality for classifying multiparticle systems}

Consider a partition of $n$-particle 
system ${\bf N}_n=\{1,2,\ldots,n\}$ into $k$
nonempty and disjoint subsets 
$\alpha_1,\ldots,\alpha_k$, where 
$\sum^k_{i=1}|\alpha_i|=n$, 
to which we refer as a $k$-partite split of the system\cite{bib:Dur2}.
Let us now consider the density operators $W$ on 
${\cal H}=\otimes_{j=1}^n{\cal H}_j$, where ${\cal H}_{j}$ represents
the Hilbert space with respect to particle $j$.

A density operator $W$ may be called $k$ separable with respect to a 
partition $\alpha_1,\ldots,\alpha_k$ iff
it can be written as
\begin{eqnarray}
W=\sum_{l} p_l 
\left(\otimes^k_{i=1}W_l^{\alpha_i}\right), \left(p_l\geq 0,\sum_l p_l=1\right)
\label{ksep},
\end{eqnarray}
where $W^{\alpha_i}_l,\forall l$ are the density operators on the partial Hilbert space 
$\otimes_{j\in\alpha_i}{\cal H}_{j}$.

Considering any subset $\alpha\subset {\bf N}_n$ and the
positive operators $X$ acting on ${\cal H}$,
let $X^{T_{\alpha}}$ 
denote the partial transpose of all sites belonging to $\alpha$.
Let ${\cal P}$ denote a family of sets, which consists of all unions of 
$\alpha_1,\ldots,\alpha_k$ together with the empty set, so that
${\cal P}$ has $2^k$ elements.
A positive operator $X$ 
may be called $k$-positive partial transpose ($k$-PPT) 
with respect to this specific partition iff
$X^{T_{\alpha}}\geq 0$ for all 
$\alpha\in{\cal P}$.

Clearly if a density operator $W$ is not $k$-PPT 
with respect to a specific partition, the state $W$
should not be $k$ separable with respect to the specific partition.

For each particle $j$, we assume there are $d_j$ orthonormal 
states as a certain basis set each, which are 
denoted by $\{|b^j_1\rangle,|b^j_2\rangle,\ldots,|b^j_{d_j}\rangle\},
j\in{\bf N}_n$,
when we represent $W$ in the matrix form.
We choose two arbitrary orthonormal states, 
which are included in each basis of all particles.
We denote the two states as $\{|0\rangle,|1\rangle\}$.
That is to say, 
\begin{eqnarray}
|0\rangle,|1\rangle\in
\{|b^j_1\rangle,|b^j_2\rangle,\ldots,|b^j_{d_j}\rangle\}, 
\forall j\in{\bf N}_n .\label{basis}
\end{eqnarray}
Let us consider a subspace ${\cal H}^2$, so that ${\cal H}^2$ is a subspace
supported by $\{|0\rangle,|1\rangle\}^{\otimes n}$, and hence 
${\rm Dim}({\cal H}^2)=2^n$, where ${\rm Dim}({\cal H})=d=\prod^n_{j=1} d_j$.

From the argument discussed by D\"ur {\it et al}. \cite{bib:Dur1,bib:Dur2}, 
we can see that 
applying local operations, 
an arbitrary positive operator of $n$ spin-1/2 systems can be transformed into one of the members of a family of positive operators as
\begin{eqnarray}
\rho_n&=&\sum_{\sigma=\pm}\lambda_0^{\sigma}
|\Psi_0^{\sigma}\rangle\langle\Psi_0^{\sigma}|\nonumber\\
&&+\sum_{j=1}^{2^{(n-1)}-1}\lambda_j
(|\Psi_j^{+}\rangle\langle\Psi_j^{+}|
+|\Psi_j^{-}\rangle\langle\Psi_j^{-}|),
\end{eqnarray}
where $|\Psi_j^{\pm}\rangle$ represent 
the orthonormal GHZ basis \cite{bib:GHZ} by
\begin{eqnarray}
&&|\Psi_j^{\pm}\rangle=
\frac{1}{\sqrt{2}}(|j\rangle|0\rangle\pm 
|2^{n-1}-j-1\rangle|1\rangle),
\end{eqnarray}
where $j=j_1 j_2 \cdots j_{n-1}$ is understood in binary notation.

If the suitable 
local unitary transformations proposed by 
D\"ur {\it et al}. \cite{bib:Dur1,bib:Dur2}, 
by which are denoted $\{U_{\zeta}\}$,
are applied on a density operator $W$, i.e., 
\begin{eqnarray}
W\rightarrow\sum_{\zeta} 
p_{\zeta}(U_{\zeta}\oplus I)W(U_{\zeta}^{^\dagger}\oplus I),
\end{eqnarray}
where $I$ 
represents the identity operator for the $(d-2^n)$-dimensional space, 
then a suboperator of 
the density operator $W$ on the subspace ${\cal H}^2$ 
can be transformed into the positive operator $\rho_n$, where 
the normalization condition ${\rm Tr}[W]=1$ leads to
${\rm Tr}_{{\cal H}^2}[\rho_n]\leq 1$.

The values of the positive coefficients of $\rho_n$, i.e., all
$\lambda$ are kept unchanged during the transformation procedure. 
Therefore,
one finds
\begin{eqnarray}
\lambda_0^{\pm}&=&{\rm Tr}[W(|\Psi^{\pm}_0\rangle\langle \Psi^{\pm}_0|
\oplus {\bf 0})],\nonumber\\
2\lambda_j&=&
{\rm Tr}[W\{(|\Psi^{+}_j\rangle\langle \Psi^{+}_j|
+|\Psi^{-}_j\rangle\langle \Psi^{-}_j|)
\oplus {\bf 0}\}],\label{coeff}
\end{eqnarray}
where ${\bf 0}$ represents the null operator 
for the $(d-2^n)$-dimensional space.
Let $\Delta$ be $\lambda_0^+ -\lambda_0^-$ and suppose that 
$
\Delta=\lambda_0^+ -\lambda_0^- \geq 0.
$

Let $\beta$ be a subset $\beta\subset{\bf N}_n$
and $l(\beta)$ be the integer $l_1\cdots l_n$ in binary notation
with $l_m=1$ for $m\in \beta$ and $l_m=0$ otherwise, and 
let $j(\beta)$ be the integer binary-represented by 
$l_1\cdots l_{n-1}$.
We define a function
\begin{eqnarray}
g:\beta\mapsto g(\beta)\in 
\{0\}\cup {\bf N}_{2^{(n-1)}-1},
\end{eqnarray}
as follows:
\begin{eqnarray}
g(\beta)=
\left\{
\begin{array}{cl}
\displaystyle
j(\beta),
&\quad l(\beta)\equiv 0\ ({\rm mod}\ 2),\\
\displaystyle
2^{n-1}-j(\beta)-1,
&\quad l(\beta)\equiv 1\ ({\rm mod}\ 2).
\end{array} \right.
\end{eqnarray}
The function $g$ is two-to-one.

From the result obtained by D\"ur {\it et al}. \cite{bib:Dur1,bib:Dur2}, 
it is easy to see that 
\begin{eqnarray}
\rho_n^{T_{\beta}}\geq 0\ {\rm iff}\ \Delta\leq 2\lambda_{g(\beta)}.
\label{Durform}
\end{eqnarray}
[We cannot change the relation (\ref{Durform}) 
even if the normalization condition with respect to 
the positive operator $\rho_n$ is modified as we have mentioned above.]
Hence, a positive operator $\rho_n$ is $k$-PPT 
with respect to a specific partition iff 
$\Delta\leq 2\lambda_{g(\alpha)}$ for all 
$\alpha\in{\cal P}$.
Even though ${\cal P}$ has $2^k$ elements,
we can see that the number of the really necessary elements of ${\cal P}$ 
is $2^{k-1}-1$ because it does not matter whether we transpose $\alpha$ or its complement and ${\cal P}$ contains the empty set.
We now define a set of integers $\tau$ by means of the two-to-one 
function $g$ as follows.
Let $\tau'$ be a set of integers $\{g(\alpha)|\alpha\in {\cal P}\}$, i.e., the image of
the family of sets ${\cal P}$, and define $\tau$ by a difference set $\tau'\backslash\{0\}$ so that $|\tau|=2^{(k-1)}-1$.
At this stage, we can show the following.

{\it Proposition}. Let $W$ be the density operators on 
${\cal H}=\otimes_{j=1}^n{\cal H}_j$ and let 
 $\hat{\Upsilon}(k)$ be an operator
$(|\Psi^{+}_0\rangle\langle \Psi^{+}_0|
-(1-2^{2-k})|\Psi^{-}_0\rangle\langle \Psi^{-}_0|)
\oplus {\bf 0}$.
Under the conditions that ${\rm Tr}[W\{(|\Psi^{+}_0\rangle\langle \Psi^{+}_0|
-|\Psi^{-}_0\rangle\langle \Psi^{-}_0|)
\oplus {\bf 0}\}]\geq 0$ and that
$W$ is $k$-PPT with respect to a specific partition, the maximum value of
${\rm Tr}[W\hat{\Upsilon}(k)]$ over $W$ is $2^{1-k}$.

{\it Proof.} 
Let $\rho$ be the suboperator of $W$ on the subspace ${\cal H}^2$.
If the density operator $W$ is positive, then any suboperator on 
the subspace,
which is supported by a subset of a basis set that supports the original 
Hilbert space ${\cal H}$, is positive \cite{bib:Peres3}.
Therefore, we find $W^{T}\geq 0\Rightarrow{\rho}^T\geq 0$.
Since the transformation procedure is performed by 
local operations, we cannot change the positivity of 
partial transpose, i.e., 
${\rho}^T\geq 0\Rightarrow{\rho_n}^T\geq 0$.
Hence if $W$ is $k$-PPT, then $\rho_n$ should be $k$-PPT.
So the following holds:
\begin{eqnarray}
\Delta\leq 2\lambda_{i}, \forall i\in\tau,\label{assum}
\end{eqnarray}
where $\tau$ is the set of integers $\{i_1,i_2,\ldots,i_{2^{(k-1)}-1}\}$,
which has been defined as mentioned above.
The relation (\ref{assum}) implies 
\begin{eqnarray}
\Delta\leq
2\min\{\lambda_{i_1},\lambda_{i_2},\ldots,\lambda_{i_{2^{(k-1)}-1}}\}.
\end{eqnarray}
On the other hand, according to the following relations
\begin{eqnarray}
{\rm Tr}_{{\cal H}^2}[\rho_n]\leq 1
\Leftrightarrow
2\sum_{j=1}^{2^{(n-1)}-1}\lambda_j+\lambda_0^+ +\lambda_0^-\leq 1,
\end{eqnarray}
we have
\begin{eqnarray}
2\sum_{j\in \tau}\lambda_j\leq 1-\lambda_0^+ -\lambda_0^-.
\end{eqnarray}
This leads to the following relation:
\begin{eqnarray}
2\min\{\lambda_{i_1},\lambda_{i_2},\ldots,\lambda_{i_{2^{(k-1)}-1}}\}\leq
\frac{1-\lambda_0^+ -\lambda_0^-}{2^{(k-1)}-1}.
\end{eqnarray}
Hence we get
\begin{eqnarray}
\Delta\leq \frac{1-\lambda_0^+ -\lambda_0^-}{2^{(k-1)}-1}
&\Leftrightarrow&
\Delta\leq 2^{1-k}-2^{2-k}\lambda_0^-,
\end{eqnarray}
and, from Eq.~(\ref{coeff}),
\begin{eqnarray}
{\rm Tr}[W\hat{\Upsilon}(k)]\leq 2^{1-k}.\label{finalineq}
\end{eqnarray}
The equality of the relation (\ref{finalineq}) holds when
${\rm Tr}_{{\cal H}^2}[\rho_n]=1$ and $\lambda_{i}=
\frac{1-\lambda_0^+-\lambda_0^-}{2(2^{(k-1)}-1)}\forall i\in\tau$.
Q.E.D.
 

D\"ur {\it et al}. showed \cite{bib:Dur1,bib:Dur2} that $\rho_n$ 
is $k$ separable with respect to a partition $\alpha_1,\ldots,\alpha_k$
iff $\rho_n^{T_{\alpha}}\geq 0$ for all 
$\alpha\in{\cal P}$ in the case that ${\rm Tr}_{{\cal H}^2}[\rho_n]=1$.
Hence it is easy to see that the inequality (\ref{finalineq})
and the equality of the relation (\ref{finalineq}) 
can hold when we assume that $W$ is $k$ separable with respect to the specific partition.


In the case that $\Delta=\lambda_0^+ -\lambda_0^- < 0$, we exchange 
$\lambda_0^+$ and $\lambda_0^-$. 
The above arguments then go in the same way.
We have chosen the two orthonormal states as $\{|0\rangle,|1\rangle\}$.
In fact, how to represent $W$ in the matrix 
form and to choose two orthonormal states like Eq.~(\ref{basis})
is arbitrary.
The above arguments do not change 
even if we choose two orthonormal states of any manner.
Therefore the relation (\ref{finalineq}) can be represented as 
\begin{eqnarray}
{\rm Tr}
[W\hat{\Upsilon}(k)]\leq 2^{1-k},
\forall \{|0\rangle,|1\rangle\}.
\end{eqnarray}
If the following is held:
\begin{eqnarray}
{\rm Tr}
[W\hat{\Upsilon}(k)]>2^{1-k},
\exists \{|0\rangle,|1\rangle\},\label{main}
\end{eqnarray}
we confirm that a given state $W$ cannot belong to $k'$-PPT state class, 
and $W$ cannot belong to $k'$ separable state 
class ($k'\geq k$).

For $k=2$, the operator $\hat{\Upsilon}(k)$ becomes the projection operator $|\Psi^{+}_0\rangle\langle \Psi^{+}_0|\oplus {\bf 0}$, and hence 
the relation (\ref{main}) is written by 
\begin{eqnarray}
{\rm Tr}[W(|\Psi^{+}_0\rangle\langle \Psi^{+}_0|
\oplus {\bf 0})]>1/2,
\exists \{|0\rangle,|1\rangle\}.
\end{eqnarray}
Therefore, for mixed multipartite systems
of an arbitrary dimension we obtain a sufficient condition 
for a density operator to have no positive partial 
transpose with respect to any subsystem, which is that the fidelity 
to the quantum state 
belonging to the generalized GHZ state class 
that was discussed by Cabello is larger than 1/2.


\section{Conclusion }

In conclusion, 
we have presented an inequality that classifies mixed multipartite systems
of an arbitrary dimension with respect to separability and positivity of partial transpose.
The inequality enables experimentally feasible classification 
of the observed state of mixed
multipartite systems
of an arbitrary dimension.
The inequality also implies that a sufficient condition 
for a density operator to have no positive partial 
transpose with respect to any subsystem is that the 
fidelity to a generalized GHZ state is larger 
than 1/2 for mixed multipartite systems
of an arbitrary dimension.

\acknowledgments


The author would like to thank M. Koashi and N. Imoto for valuable discussions.

\end{document}